\documentclass[twocolumn,english,prl,amssymb,aps,superscriptaddress,showpacs,twocolumn,amsmath,showkeys,floatfix]{revtex4-1}
\usepackage[T1]{fontenc}
\usepackage[latin9]{inputenc}
\usepackage[active]{srcltx}
\usepackage{textcomp}
\usepackage{amsmath}
\usepackage{graphicx}
\usepackage{color}
\usepackage{esint}
\usepackage{caption}

\makeatletter
\usepackage{babel}

\makeatother

\begin{document}

%\title{Nanolaser Mode-Locking of Hermite-Gaussian Modes}
\title {Investigation of analog signal distortion introduced by a fiber phase sensitive amplifier}
\author{Debanuj Chatterjee}
\affiliation{Universit\'e Paris-Saclay, CNRS, ENS Paris-Saclay, CentraleSup\'elec, LuMIn, Gif-sur-Yvette, France}
\author{Yousra Bouasria}
\affiliation{Equipe Sciences de la Mati\`ere et du Rayonnement (ESMaR), Faculty of Sciences, Mohammed V University, Rabat, Morocco}
\author{Weilin Xie}
\affiliation{School of Optics and Photonics, Beijing Institute of Technology, 100081 Beijing, China}
\author{Tarek Labidi}
\affiliation{Universit\'e Paris-Saclay, CNRS, ENS Paris-Saclay, CentraleSup\'elec, LuMIn, Gif-sur-Yvette, France}
\author{Fabienne Goldfarb}
\affiliation{Universit\'e Paris-Saclay, CNRS, ENS Paris-Saclay, CentraleSup\'elec, LuMIn, Gif-sur-Yvette, France}
\author{Ihsan Fsaifes}
\affiliation{Ecole Polytechnique, Institut Polytechnique de Paris, 91128 Palaiseau Cedex, France}
\author{Fabien Bretenaker}
\affiliation{Universit\'e Paris-Saclay, CNRS, ENS Paris-Saclay, CentraleSup\'elec, LuMIn, Gif-sur-Yvette, France}
\affiliation{Light and Matter Physics Group, Raman Research Institute, Bangalore 560080, India}

\begin{abstract} 
%Microwave photonic links are at the heart of modern day analog RF systems due to their wide bandwidth and low loss profile. Introduction of phase sensitive amplifiers (PSA) in such microwave photonic links can further extend the length of the links by compensating the acquired signal losses. However, the PSA can be a source of distortions, degrading the signal in the link. 
We numerically simulate the distortion of an analog signal carried in a microwave photonics link containing a phase sensitive amplifier (PSA), focusing mainly on amplitude modulation format. The numerical model is validated by comparison with experimental measurements. By using the well known two-tone test, we compare the situations in which a standard intensity modulator is used with the one where a perfectly linear modulator would be employed. We also investigate the role of gain saturation on the nonlinearity of the PSA. Finally, we establish the conditions, in which the signal nonlinearity introduced by the PSA itself can be extremely small.% we show numerically that under the non-pump depletion approximation, the PSA does not introduce extra distortions in the link. We also investigate the saturation behaviour of the link and discuss the effect of pump depletion in the PSA on the nonlinearity of the link.
\end{abstract}

%\pacs{??}

\maketitle
\section{Introduction}
Microwave photonics plays a crucial role in the development of modern day analog RF systems with applications such as antenna remoting, radio-over-fiber, phased array antennas, etc. The inclusion of optical links for the transfer of analog RF signals leads to several advantages compared to traditional electrical distribution lines such as coaxial cables or waveguides, such as an increase of processing bandwidth, decrease of signal loss, immunity to electromagnetic interference, etc.   \cite{seeds2006microwave,capmany2007microwave,yao2009microwave,yao2012tutorial}. To improve the performance of the links further, research efforts in this 
field have focused on reducing link loss, noise figure and distortion and thereby increasing the dynamic range \cite{chang2002rf,vilcot2003microwave,cox2006analog}. However, for achieving links with a longer range, optical amplifiers are necessary to boost the signal periodically. % (see Fig. \ref{linkscheme}). 
The gain of traditional optical amplifiers, such as Erbium Doped Fiber Amplifiers (EDFA) is independent of the phase of the signal. The noise figure of such phase insensitive amplifiers (PIA) cannot be smaller than 3~dB \cite{caves1982quantum,xie2014high}. On the other hand, in some fiber optic parametric amplifiers (FOPA), the signal gain depends on the relative phase between the different waves injected in the amplifier. Such a phase sensitive amplifiers (PSA)  can have a noise figure as low as 0 dB \cite{caves1982quantum,tong2011towards,labidi2018phase}. In the context of microwave photonic links, utilization of a PSA can offer the possibility of realizing other functionalities such as optical filtering through phase sensitive amplification and de-amplification \cite{agarwal2011optically}, wideband photonic assisted radio over fiber systems \cite{sodre2016photonic} and microwave photonic frequency measurement \cite{emami2014improved}.

The possibility  of noiseless amplification in PSAs has been extensively studied in the context of digital optical communication links \cite{hansryd2002fiber,tong2013low,marhic2015fiber,albuquerque2015experimental,karlsson2015transmission,olsson2018long}. Nowadays digital links are widely used across communication platforms, but such links are limited by the bandwidth of analog-to-digital converters (ADC). Although efforts to overcome such limitations were made using photonics based ADCs \cite{ghelfi2014fully}, fully analog microwave photonic links are still promising. Preliminary studies on performance of analog links with respect to modulation formats \cite{kalman1994dynamic}, signal multi-casting \cite{bres2010low} and PSA inclusion in a link \cite{lim2008ultra,agarwal2012rf,fsaifes2014intermodulation} were performed. However, the possibility of incorporating a PSA within an analog microwave photonic link needs further exploration with respect to its noise characteristics and its dependence on various parameters of modulation and PSA processes.

%\begin{figure}[H]
%\centering 
%\includegraphics[scale=0.23,trim=4 4 4 4,clip]{fig11.pdf}
%\caption{Scheme of a microwave photonic link with PSA. E/O : electric to optic; O/E : optic to electric; OFC : optical fiber cable; PSA : phase sensitive amplifier (optical amplifier).}
%\label{linkscheme}
%\end{figure}

An important factor for the performance of a microwave photonic link is the signal fidelity across the link. In a traditional microwave photonic link without PSA, the E/O (Electric to Optic) conversion process is achieved with a Mach-Zehnder modulator (MZM). This MZM, due to its nonlinear transfer function, is a source of distortion in the link. Moreover, when a PSA is added in such a link, further distortions might be generated due to nonlinear processes in the optical amplification. Generation and mitigation of nonlinearities by a PSA has been  comprehensively studied for digital links \cite{slavik2010all,liu2015all,olsson2015nonlinear,perentos2016qpsk,eliasson2016mitigation,astra2017dispersion,bottrill2017full}. But, in the case of analog links, it is still a topic of ongoing research \cite{bhatia2014linearization}. An important question in this context is to investigate the nonlinearities coming from the PSA process only. While it is well known that such distortions are negligible in the case of an EDFA \cite{Desurvire2002}, the question still deserves to be investigated in the case of a PSA. In this article, we thus use numerical modelling to find nonlinearities that are generated solely due to the addition of a PSA and not from the nonlinear MZM of the link. Since PSAs based on two pumps and degenerate signal and idler exhibit cascaded four-wave mixing phenomena that make them difficult to model \cite{baillot2014multiple,xie2015investigation}, we focus here on a PSA scheme with  degenerate pumps and nondegenerate signal and idler. The strength of the distortions induced on the signal is characterized using the so-called two-tone test in which the carrier is modulated by two nearby frequencies from which the system nonlinearities generate third order inter-modulation products (IMD3). The amplitude of these IMD3's is used as a measure of the nonlinearity of the PSA. Moreover, under certain circumstances, like high optical power of the signal or longer length of the fiber optic amplifier, the PSA gain can attain saturation and hence can change several properties of the link \cite{cristofori2011saturation,lundstrom2012phase,cristofori2013experimental}. We thus also investigate the distortion performance of the link when the PSA gain reaches saturation.

%An effective way to characterize the strength of distortions in a link is to perform the two-tone test. This test requires the generation of third order inter-modulation products (IMD3) by modulating a carrier signal with two RF tones. Next, an observation of the output RF power of these IMD3 waves is used as a measure of how noisy the link is. In principle, the output RF power of the IMD3 waves can depend on several parameters in the link related to the modulation and the PSA process when a PSA is added in the link. In this article we numerically investigate the dependence of the output RF power of the IMD3 waves on two important parameters, i.e. modulation power of the modulator and input signal power of the optical carrier. In general, a PSA is employed in two common schemes, (a) degenerate signal and idler and non-degenerate pump and (b) degenerate pump and non-degenerate signal and idler. In this analysis, we consider a PSA scheme with a degenerate pump and nondegenerate signal and idler because in a non-degenerate pump scheme, the four wave mixing between the two strong pumps lead to generation of very strong sidebands\cite{baillot2014multiple,xie2015investigation} which makes the system difficult to analyze. 

The paper is organized as follows. In Section \ref{model} we introduce the theoretical model that we use to numerically simulate the microwave photonic link with a PSA, and to predict the strength of the IMD3. Then in Section \ref{experiment} we validate our model by comparison with experimental results. In Section \ref{saturationRM} and \ref{saturationIM} we study the saturation behaviour of the link with respect to two parameters: a) input modulation power and b) input signal power, for a standard (Section \ref{saturationRM}) and a perfectly linear (Section \ref{saturationIM}) intensity modulator. In both cases, we also evaluate the spurious free dynamic range (SFDR) that can be expected from a microwave photonics link containing such a PSA. In Section \ref{discussion} we discuss the results and project some future perspectives.

\section{Model}\label{model}
\subsection{Nonlinear Schr\"odinger Equation (NLSE)}
We consider here the amplification of an optical signal in a highly nonlinear fiber (HNLF). Such fibers with a non-zero $\chi^{(3)}$ susceptibility lead to processes like self phase modulation, cross phase modulation and four-wave mixing (FWM) when different wavelengths propagate through it. This latter process can lead to transfer of energy between these different waves. In a degenerate pump PSA scheme, a strong pump and  weak signal and idler fields are launched into the HNLF. Under appropriate phase matching conditions and relative phase value, pump photons get converted into signal and idler photons leading to optical amplification of the signal and idler fields. When only these three waves need to be considered, a standard 3-wave model suffices to describe the field evolution in the HNLF. However, for larger input signal powers or for modulated signal and/or idler, the number of frequencies that need to be considered increases significantly \cite{baillot2014multiple,xie2015investigation,qian2017investigation,pakarzadeh2018two}. Since in our case the modulated signal and idler fields contain many sidebands, and the  signal powers may be large, we require a more robust model to analyze the evolution of different waves through the HNLF. Hence we resort to the Nonlinear Schr\"odinger equation (NLSE), which describes the evolution of the total field as a whole. It is given by \cite{agrawal2013}:

\begin{equation}\label{nlse4}
        \frac{\partial A}{\partial z}+\frac{i\beta^{(2)}}{2}\frac{\partial^2 A}{\partial T^2}-\frac{\beta^{(3)}}{6}\frac{\partial^3 A}{\partial T^3}-\frac{i\beta^{(4)}}{24}\frac{\partial^4 A}{\partial T^4}+\frac{\alpha}{2}A-i\gamma |A|^2A=0\ ,
\end{equation}
where $A(z,T)$ is the complex amplitude of the slowly varying envelope of the field propagating along the $z$ direction through the HNLF. As we consider that all the waves are co-polarized, $A$ can be treated as a scalar. This equation is written in a reference frame that is moving at the group velocity. Thus $T=t-\beta^{(1)}z$ where $t$ is the time in a static reference frame. The derivatives $\beta^{(n)}$ are given by $\frac{d^n\beta}{d\omega^n}$ at $\omega=\omega_c$, where $\beta(\omega)$ is the propagation constant of a wave with angular frequency $\omega$ and $\omega_c$ is a chosen central frequency, which for our case is the angular frequency of the pump. The quantity $\gamma$ is the nonlinear coefficient of the HNLF and $\alpha$ its attenuation coefficient.  

\subsection{Algorithm}
In order to numerically simulate the propagation of waves through the HNLF, we use the standard split step Fourier method (SSFM) algorithm \cite{agrawal2013} to solve the NLSE in MATLAB. In this algorithm, the dispersion and nonlinear effects of the HNLF are treated separately \cite{fisher1975numerical}.
First we generate the electric field of the input wave which consists of the degenerate pump and modulated signal and idler fields. % In the case of a perfectly linear intensity modulator, we keep only the first order modulation sidebands of the signal and idler. 
Then we launch it through the HNLF which is divided into small segments in which dispersion and nonlinearity are treated separately \cite{sinkin2003optimization}. %Each segment is further divided into two halves, one to handle dispersion in the frerquency . For propagation through each segment, first we consider the field in the frequency domain and propagate it through the first half of the segment considering only dispersion and without considering any nonlinearity of the medium. Then we move to the time domain and add the nonlinearity due to the whole segment at once without considering any dispersion of the fiber. Then we again transform the field to the frequency domain and dispersively propagate it through the second half of the segment without any nonlinearity, to get the output field at the end of a segment. Although in a real fiber, the effects of dispersion and nonlinearity act simultaneously, nevertheless with small size of the segments, this technique of treating dispersion and nonlinearity separately can be substantially accurate and numerically efficient\cite{sinkin2003optimization}. 
After the HNLF, we remove the pump and the idler parts of the spectrum keeping only the signal and its sidebands. Then we calculate the absolute square of the complex field with proper scaling to get the voltage detected by a photodiode. Finally we retrieve the RF power corresponding to different RF frequencies from the spectrum as would be obtained from an electrical spectrum analyzer (ESA).

\subsection{PSA Gain}
In the case of a PSA with a degenerate pump configuration and under the non-pump depletion approximation, the optical gain of the signal and  idler $G_{PSA}$, can be determined from the 3-wave model and is given by \cite{agrawal2013}:
\begin{equation}
\label{gaineq}
    \begin{split}
       G_{PSA}=1+\Bigg(1+&\frac{\kappa^2+4\gamma^2{P_P}^2+4\kappa\gamma P_P\cos(\Theta)}{4g^2}\Bigg)\sinh^2(gL)\\
       &+\frac{\gamma P_P  \sin(\Theta)}{g}\sinh(2gL)\ ,
    \end{split}
\end{equation}
where we have supposed that the input signal and idler have equal powers and where $\gamma$ is the nonlinear coefficient of the HNLF, $P_P$ is the input pump power, $L$ is the total length of the HNLF, $\kappa=2\gamma P_P+\Delta\beta$,  $\Delta \beta$ is the linear phase mismatch between the interacting waves and $g^2={(\gamma P_P)}^2-{(\kappa/2)}^2$. The relative phase difference $\Theta$ is given by 
\begin{equation}\label{eq03}
\Theta=\theta_s+\theta_i-2\theta_P\ ,
\end{equation}
where $\theta_s$, $\theta_P$, $\theta_i$ are the phases of the signal, pump and the idler respectively at the input of the HNLF.

From Eq.\,(\ref{gaineq}), we see that with the other parameters being constant, the PSA gain depends on the relative phase $\Theta$ between the pump, signal, and idler fields in a sinusoidal manner .

\subsection{Amplitude Modulation (AM) vs Phase Modulation (PM)}
The electro-optic conversion of RF signals in a microwave photonic link is performed by a modulator. The most common modulation formats are amplitude modulation (AM), which is most often performed by a Mach-Zehnder modulator (MZM), and phase modulation (PM), generally achieved using a single waveguide created in an electro-optic material. 

\begin{figure}[htp]
\centering
\includegraphics[width=0.9\columnwidth]{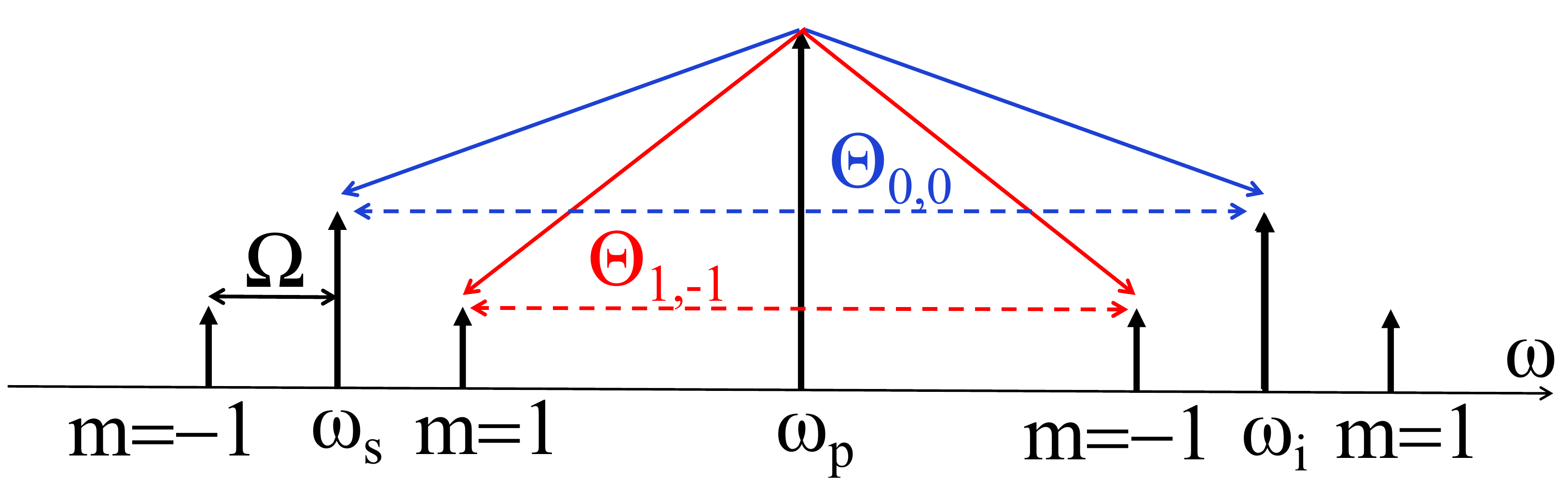} \caption{Signal ($\omega_s$), idler ($\omega_i$), and pump ($\omega_p$) spectra with their first order sidebands shifted by $\pm\, \Omega$. $\Theta_{m,n}$ represents the relative phase between the pump and the waves at frequencies $\omega_s+m\Omega$ and $\omega_i+n\Omega$ (not to scale).}\label{ampmfig}
\end{figure}

The unmodulated signal or idler electric field $E_{j}(t)$ at the input of the modulator reads:
\begin{equation}
    E_{j}(t)=\mathcal{E}_{j} e^{-i\omega_{j} t}+c.c. \;.
\end{equation}
where $j=s,i$ for signal or idler, respectively. $\omega_j$ is the angular frequency and $\mathcal{E}_{j}$ the complex amplitude of the field, and $c.c.$ means complex conjugate. For modulation with a single frequency $\Omega$, the applied voltage $V(t)$ is given by :
\begin{equation}
    V(t)=V_{DC}+V_{AC}\cos(\Omega t+\varphi)\ ,
\end{equation}
where $V_{DC}$ is the DC bias of the modulator, $V_{AC}$ is the amplitude of the sinusoidal modulation and $\varphi$ is the initial phase of the modulation signal.

For AM, the output field of a MZM operating in push-pull configuration is given by :    \begin{equation}\label{ameq}\small
\begin{split}E_{out,j}(t)&=\frac{\mathcal{E}_{j}}{2}e^{-i\omega_{j} t}\left[e^{i\phi}\sum_{m=-\infty}^{\infty}i^mJ_m(\zeta){e^{im\Omega t}}e^{im\varphi}\right.\\
&\left.+e^{-i\phi}\sum_{p=-\infty}^{\infty}i^pJ_p(-\zeta){e^{ip\Omega t}}e^{ip\varphi}\right]+c.c.\;,\end{split}\end{equation}
where $\zeta=\frac{\pi V_{AC}}{V_{\pi}}$, $\phi=\frac{\pi V_{DC}}{ V_{\pi}}$, $V_{\pi}$ is the half-wave voltage of the MZM, $m$ and $p$ are integers and $J_n$ represents the Bessel function of the first kind of order $n$. 

In the case of PM, we take $V_{DC}$ equal to 0 and the output signal (or idler) electric field after the phase modulator is given by : 
\begin{equation}\label{pmeq}
\begin{split}
    &E_{out,j}(t)=\mathcal{E}_{j} e^{-i\omega_{j} t} \sum_{m=-\infty}^{\infty}i^mJ_m(\zeta){e^{im\Omega t}}e^{im\varphi}+c.c.\;\;.
    \end{split}
\end{equation}
%The input relative phase $\theta_{m}$ of the sideband at angular frequencies $\omega_{j}+m\Omega$ with respect to the signal (or idler) carrier at the input of the PSA, can be determined from Eq.\,(\ref{ameq}) and Eq.\, (\ref{pmeq}) for AM and PM respectively, as shown in Table \ref{tab}.

If we consider a particular signal sideband at frequency $\omega_s-m\Omega$, the PSA gain of this sideband will also be phase dependent. If we call $\theta_{s,m}$ the phase of this sideband and  $\theta_{i,n}$ the phase of the idler sideband at frequency $\omega_i-n\Omega$, then the gain of the signal sideband considered above depends on the relative phase $\Theta_{m,-m}$ where we have noted, similarly to Eq.\,(\ref{eq03}):
\begin{equation}
\Theta_{m,n}=\theta_{s,m}+\theta_{i,n}-2\theta_p\ .\label{eq08}
\end{equation}

\begin{center}
\begin{tabular}{|p{0.8cm}|p{2cm}|p{1.5cm}|p{1.5cm}| }
\hline
$m$& Frequency&$\theta_{j,m}$ (AM) & $\theta_{j,m}$ (PM) \\
\hline
1&$\omega_{j}-\Omega$ & $-\pi+\varphi$& 
$\frac{\pi}{2}+\varphi$ \\
\hline
-1&$\omega_{j}+\Omega$&$-\pi-\varphi$&$\frac{\pi}{2}-\varphi$\\
\hline
\end{tabular}
\captionof{table}{Expressions of $\theta_{j,m}$ for $m=1$ and $m=-1$ in the cases of AM and PM.}\label{tab}
\end{center}

%In the PSA, the gain of a given signal modulation sideband at frequency $\omega_s+m\Omega$ depends on the relative phase between itself, the pump, and the symmetrically opposite sideband of the idler, at frequency $\omega_i-m\Omega$, as shown in Fig.\,\ref{ampmfig}. We denote as  $\Theta_{m,n}$ the relative phase for the FWM process involving the pump and the waves at frequencies $\omega_s+m\Omega$ and $\omega_i+n\Omega$ , where $m$ and $n$ are integers, as shown in Fig.\,\ref{ampmfig}.
Focusing on the first sidebands ($m=\pm1$), let us consequently use Eqs.\,(\ref{ameq}) and (\ref{pmeq}) to derive the phase of these sidebands with respect to the signal carrier in the cases of AM and PM. First, in the case of AM, taking the terms corresponding to $m=1$ in Eq.\,(\ref{ameq}) leads to the following expression for the complex field of the $m=1$ sideband:
\begin{equation}
\mathcal{E}_{j}\,e^{-i\omega_jt}J_1(\zeta)e^{i\Omega t}\sin\phi\, e^{i(\varphi-\pi)}\ ,
\end{equation}
showing that $\theta_{j,1}=-\pi+\varphi$, as given in Table \ref{tab}. Similarly, taking the terms corresponding to $m=-1$ in Eq.\,(\ref{ameq}) leads to $\theta_{j,-1}=-\pi-\varphi$, as reported in the last line of Table \ref{tab}.
These expressions lead to the following relation between the relative phase $\Theta_{1,-1}$ for the first-order sidebands and the relative phase $\Theta_{0,0}$ for the signal and idler carriers:
\begin{equation}\label{relpham}
    \Theta_{1,-1}^{AM}=\Theta_{0,0}-2\pi\ .
\end{equation}
This relation shows that when, in the case of AM,  the gain for the signal and idler carriers is maximum then the gain for the first-order sidebands is also maximum. This guarantees an optimal amplification of the AM signal by the PSA.

%of  leads to the following relation between the relative phase $\Theta_{0,0}$ for FWM between the pump, signal, and idler and the relative phase $\Theta_{1,-1}$ between the pump and the first symmetric sidebands of the signal and idler:
%\begin{equation}\label{relpham}
%    \Theta_{1,-1}=\Theta_{0,0}-2\pi=\Theta_{max}%.
%\end{equation}
%This leads to the fact that when $\Theta_{0,0}=\Theta_{\mathrm{max}}$, i. e., when the gain for the signal and idler carriers is maximum, then $\Theta_{1,-1}$ is also equal to $\Theta_{\mathrm{max}}$ and these sidebands also experience the maximum amplification. 

%When the relative phase $\Theta_{0,0}$ for the PSA process involving the signal, the pump, and the idler is such that the PSA gain of the signal is maximum, i.e. $\Theta_{0,0}=\Theta_{max}$ then for AM, the relative phase for the PSA process with the waves at frequencies $\omega_s+\Omega$, $\omega_p$ and $\omega_i-\Omega$ is also maximized for their relative phase $\Theta_{1,-1}$ to be $\Theta_{max}$, given the modulation frequency $\Omega$ is small. For such a process, relative phase $\Theta_{1,-1}$ is obtained using Table \ref{tab} :
%
%\begin{equation}\label{relpham}
 %   \Theta_{1,-1}=\Theta_{max}-2\pi=\Theta_{max}.
%\end{equation}
%
%Thus the maximum PSA gain for the signal will also lead to maximum PSA gain for the first order sideband in the case of AM. 
In contrast, in the case of PM, Eq.\,(\ref{pmeq}) leads to the following expression for the complex field of the $m=1$ sideband:
\begin{equation}
\mathcal{E}_{j}\,e^{-i\omega_jt}J_1(\zeta)e^{i\Omega t}\,e^{i(\pi/2+\varphi)}\ ,
\end{equation}
leading to $\theta_{j,1}=\pi/2+\varphi$ as given in the last column of Table \ref{tab}. Similarly, $m=-1$ leads $\theta_{j,-1}=\pi/2-\varphi$. Thus, in the case of PM, Eq.\,(\ref{relpham}) becomes:
\begin{equation}\label{relphpm}
    \Theta_{1,-1}^{PM}=\Theta_{0,0}+\pi\ .
\end{equation}

%Table \ref{tab} shows
%that Eq.\,(\ref{relpham}) must be replaced by the following relation:%we see that when the signal PSA gain is maximized, i.e $\Theta_{0,0}=\Theta_{max}$, the corresponding value of $\Theta_{1,-1}$ is given by :
%\begin{equation}\label{relphpm}
%    \Theta_{1,-1}=\Theta_{0,0}+\pi.
%\end{equation}
This equation shows that for PM format, a maximum PSA gain for the signal and idler carriers leads to de-amplification of the first order sidebands, and vice-versa. Since at the output of the HNLF, the strength of the output RF signal mainly originates from the beating between the optical carrier and the first order sidebands, one can see that a PSA is a good choice to amplify AM signals but not for PM signals. We thus consider only AM in the following. %while an AM thus for AM the output RF signal will be amplified while for PM, it will be much weaker due to the negative gain of the first order sidebanbs. 

\begin{figure}[htp]
\centering
\includegraphics[width=\columnwidth]{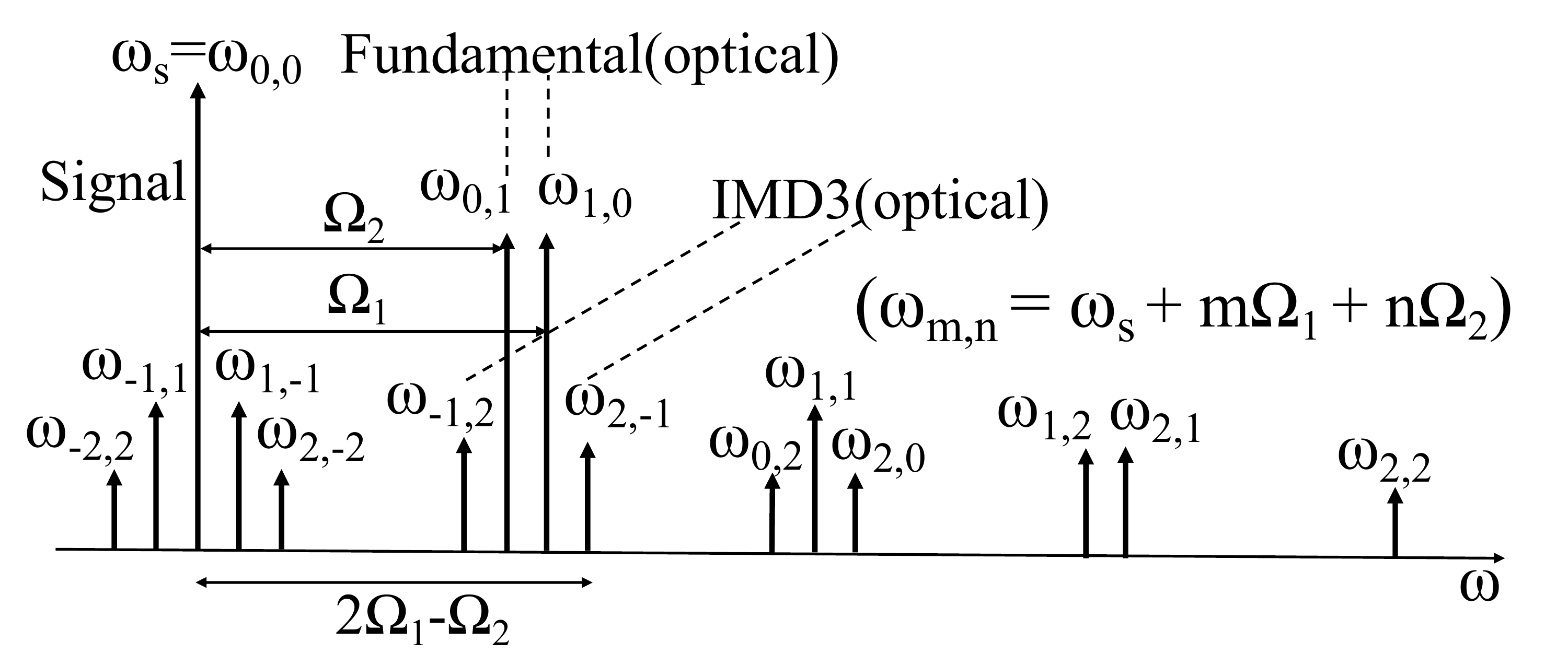}
\caption{Optical waves generated after modulation of the signal with a MZM. Here $\omega_{m,n}=\omega_s+m\Omega_1+n\Omega_2$ with $m$ and $n$  arbitrary integers and $\omega_s$ the signal frequency (not to scale). }\label{spectrumrealmzm}
\end{figure}
\subsection{Two-Tone Test}\label{sec2.E}
The two-tone test is a good indicator of the linearity of a component or a link. It consists of launching two RF frequencies $\Omega_1$ and $\Omega_2$ with equal amplitudes and looking at the amplitudes of the third-order intermodulation products (IMD3's) at frequencies $2\,\Omega_1-\Omega_2$ and $2\,\Omega_2-\Omega_1$ at the output of the device under test. We thus suppose that we perform AM of a carrier at frequency $\omega_{j}$ ($j=s,i$) using the following modulation signal:

%
%, are fed into a modulator that modulates an optical carrier (signal and/or idler) at frequency $\omega_{j}$ (where subscript $j=s$ is for signal and $j=i$ for idler). Similar to the previous section, the modulation signal is given by :
\begin{equation}
V(t)=V_{DC}+V_{AC}\left(\cos\Omega_1 t+\cos\Omega_2 t\right)\ .
\end{equation}
Then, the field at the output of a MZM operating in push-pull configuration becomes:
\begin{equation}\label{mzmeouteq}\small
\begin{split}E_{out,j}(t)
&=\frac{\mathcal{E}_{j}}{2}e^{-i\omega_j t}\Bigg[e^{i\phi}\left(\sum_{m=-\infty}^{\infty}i^mJ_m(\zeta){e^{im\Omega_1 t}}\right)\\ \times &\left(\sum_{n=-\infty}^{\infty}i^nJ_n(\zeta){e^{in\Omega_2 t}}\right)\\
&+e^{-i\phi}\left(\sum_{p=-\infty}^{\infty}i^pJ_p(-\zeta){e^{ip\Omega_1 t}}\right)\\ \times &\left(\sum_{q=-\infty}^{\infty}i^qJ_q(-\zeta){e^{iq\Omega_2 t}}\right)\Bigg]+c.c.\end{split}\end{equation}
We can see from Eq.\,(\ref{mzmeouteq}) that, apart from the first order modulated sidebands (that we will name optical fundamental in the following) at optical frequencies $\omega_{j} \pm \Omega_{1}$ and $\omega_{j} \pm \Omega_{2}$, many extra tones at frequencies $\omega_{j} \pm m\Omega_{1}\pm n\Omega_{2}$ are generated in the output signal, where $m$ and $n$ are arbitrary integers (see Fig.\,\ref{spectrumrealmzm}). This is due to the nonlinear transfer function of the MZM. The tones generated at frequencies $\omega_{j} \pm 2\Omega_{1} \mp \Omega_2$ and $\omega_{j} \pm 2\Omega_{2} \mp \Omega_1$ are known as the optical third order intermodulation products (called optical IMD3).
%\begin{figure}[h]
%\centering
%\includegraphics[scale=0.3]{spectrum1.pdf}
%\includegraphics[scale=0.3]{spectrum2.png}
%\caption{Optical waves generated after modulation of the signal with a MZM. Here $\omega_{m,n}=\omega_s+m\Omega_1+n\Omega_2$ with $m$ and $n$  arbitrary integers and $\omega_s$ the signal frequency (not to scale). }\label{spectrumrealmzm}
%\end{figure}

\begin{figure}[htp]
\centering
\includegraphics[scale=0.25]{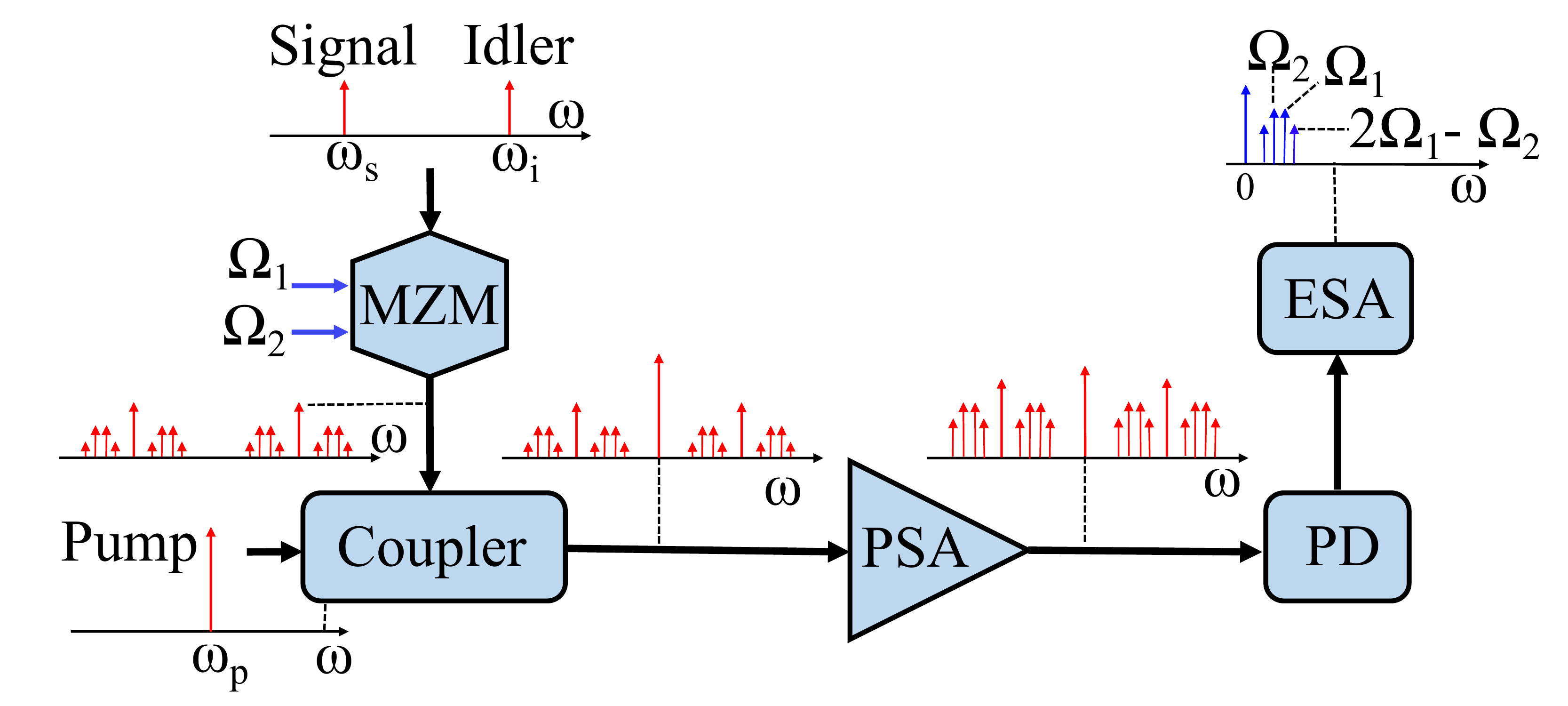}
\caption{Principle of two-tone test of a microwave photonic link including a PSA. The signal ($\omega_s$) and idler ($\omega_i$) waves are modulated at two  frequencies $\Omega_1$ and $\Omega_2$ and combined with a pump ($\omega_p$) wave and launched in a HNLF (working as PSA). The spectra shown in red (blue) correspond to optical (RF)  frequencies. MZM : Mach-Zehnder modulator; PD : Photodetector; ESA : Electrical spectrum analyzer. }\label{scheme1}
\end{figure}

When $|\Omega_1-\Omega_2|$ is small, the optical IMD3 waves can be very close in frequency to the optical fundamental waves and hence can be a significant source of distortion very difficult to filter out. Also, any component inserted within the link, such as a PSA, can further amplify the optical IMD3 waves and degrade the link. At the end of the link, a photodiode (PD) is used for detection of signals. It down-converts the beating between the different optical waves (optical carrier, optical fundamental, optical IMD3, etc.) and into a RF signal. The generated output current $I_{PD}$ is thus given by :
\begin{equation}
    I_{PD}=sP_{opt}.
\end{equation}
where $s$ is the sensitivity of the detector and $P_{opt}$ is the total optical power incident on the detector, which can be calculated by integrating the norm of the time-averaged Poynting vector of the optical wave over the beam cross-section. The beating of the optical carrier with the optical fundamental (or optical IMD3) generates the RF fundamental (or RF IMD3) signal at the output. This RF signal can be fed into an electrical spectrum analyzer (ESA) to determine the output powers of the RF fundamental and IMD3 tones. The dependence of the output power of the RF IMD3 signal on the input modulation power of the link can be used to characterize its linearity performance \cite{kalman1994dynamic,cox2006analog}. A scheme for such an experiment is presented in Fig.\,\ref{scheme1}.

%\begin{figure}[h]
%\centering
%\includegraphics[scale=0.25]{scheme41.pdf}
%\includegraphics[scale=0.3]{spectrum2.png}
%\caption{Principle of two-tone test of a microwave photonic link including a PSA. The signal ($\omega_s$) and idler ($\omega_i$) waves are modulated at two  frequencies $\Omega_1$ and $\Omega_2$ and combined with a pump ($\omega_p$) wave and launched in a HNLF (working as PSA). The spectra shown in red (blue) correspond to optical (RF)  frequencies. MZM : Mach-Zehnder modulator; PD : Photodetector; ESA : Electrical spectrum analyzer. }\label{scheme1}
%\end{figure}

\begin{figure*}[htp]
\centering
\includegraphics[width=0.9\textwidth]{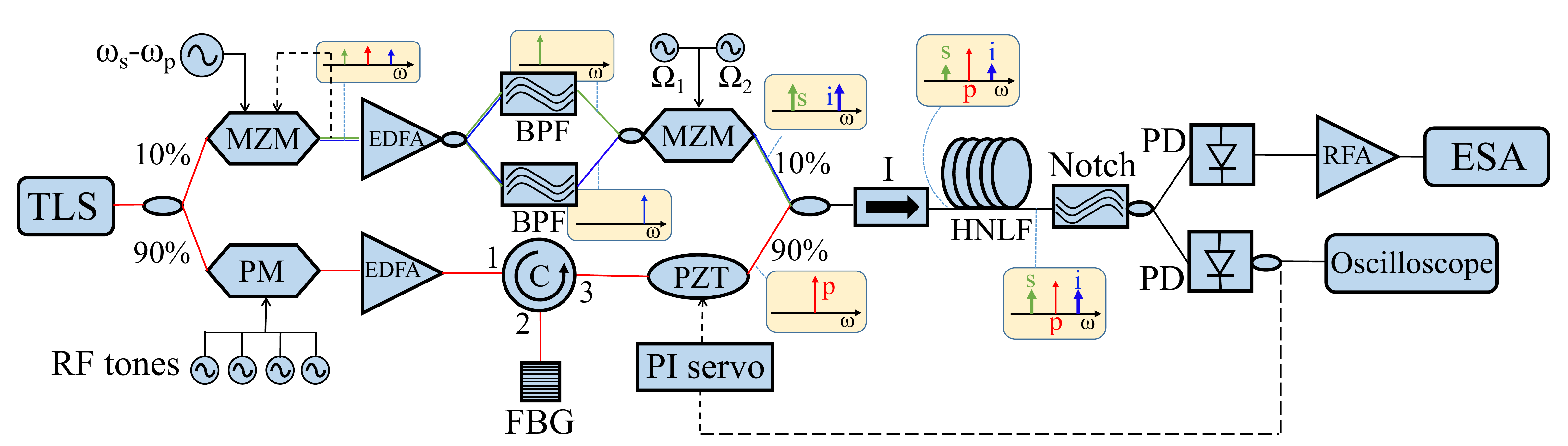}
\caption{Experimental setup for the two tone test in a microwave photonic link with PSA. The spectrum at different positions of the setup are shown in yellow boxes (not to scale) where s, i and p represent signal, idler and pump waves. The feedback loops are shown in dashed arrows. TLS : Tunable laser source; MZM : Mach-Zehnder modulator; PM : Phase modulator; EDFA : Erbium doped fiber amplifier; C : Circulator; BPF : Bandpass filter; FBG : Fiber Bragg grating; PZT : Piezo-electric transducer; I : Isolator; HNLF : Highly nonlinear fiber; Notch : Notch filter; PD : Photodetector; RFA : RF amplifier; ESA : Electrical spectrum analyzer.}\label{exptsetup}
\end{figure*}

\section{Experimental Validation of the Model}\label{experiment}
In order to validate our numerical model, we compare our numerical results with experimental data. We developed a test setup mimicking a microwave photonic link including a PSA following the scheme in Fig.\,\ref{scheme1}. A detailed scheme of the setup is shown in Fig.\,\ref{exptsetup}. The output of a tunable laser source (TLS) is divided into two branches. On one branch, the light is modulated by a RF signal at frequency $(\omega_s-\omega_p)/2\pi=20\,\mathrm{GHz}$ to create the signal and idler from the laser source tuned at a central frequency of 1547\,nm. It is then amplified with an EDFA and filtered with two bandpass filters (BPF) tuned at the signal and idler frequencies to get rid of the central frequency. Then the signal and idler are modulated with two RF tones ($\Omega_1$ and $\Omega_2$) using a MZM in push-pull configuration. The other branch of the TLS is fed with several RF tones to suppress stimulatted Brillouin scattering (SBS). Then it is amplified with an EDFA. A fiber Bragg grating (FBG) with 0.2 nm optical bandwidth at 1547 nm along with a circulator (C) are used to filter out the amplified spontaneous emission (ASE) pedestal on the pump spectrum coming from the EDFA. A piezo-electric transducer (PZT) is used to control the phase of the pump. Then the two branches are coupled and fed into a HNLF which acts as a PSA. An isolator is used at the entry of the HNLF to eliminate any unwanted back reflection. As the PSA gain depends on the relative phase between the pump, signal, and idler, the relative phase is servo-locked such that the PSA gain of the signal is maximized. At the exit of the HNLF, the pump and the idler are removed using a notch filter. One part of this signal is detected by a photodiode and fed back using a PI servo loop to the PZT to control the phase of the pump. The other part is detected by a photodiode and is fed into an electrical spectrum analyzer (ESA) for retrieving the power of the RF fundamental and IMD3 waves for different input modulation powers.

The different parameters used in the experiment are the following. The pump wavelength is 1547 nm. The pump-signal frequency separation is $(\omega_s-\omega_p)/2\pi=20\,\mathrm{GHz}$. The modulation frequencies $\Omega_1/2\pi$ and $\Omega_2/2\pi$ are 1\,GHz and 0.998\,GHz, respectively. The half-wave voltage $V_{\pi}$ of the modulator is 5 V and the DC bias $V_{DC}$ is adjusted at 2.5 V. The pump power at the entry of HNLF is 22 dBm and the input signal and idler powers, including their sidebands, are equal to -16 dBm. The 1000-m-long HNLF has a nonlinear coefficient $\gamma= 11.3\, \mathrm{(W.km)}^{-1}$, a zero-dispersion wavelength $\lambda_{ZDW}=1547\,\mathrm{nm}$, an attenuation coefficient $\alpha$ corresponding to 0.9 dB/km, and a dispersion slope equal to 0.017\,ps.nm$^{-2}$.km$^{-1}$. 

In the experiment, the output fundamental and IMD3 RF powers are recorded as a function of the input RF power applied to the modulator. The parameters used in the simulations are extracted from the experiment. We only changed the RF frequencies $\Omega_1/2\pi$ and $\Omega_2/2\pi$ to 1.567\,GHz and 1.564\,GHz, respectively, in order to avoid spectral leakage errors and computational complexities.

\begin{figure}[htp]
\centering
\includegraphics[width=0.7\columnwidth]{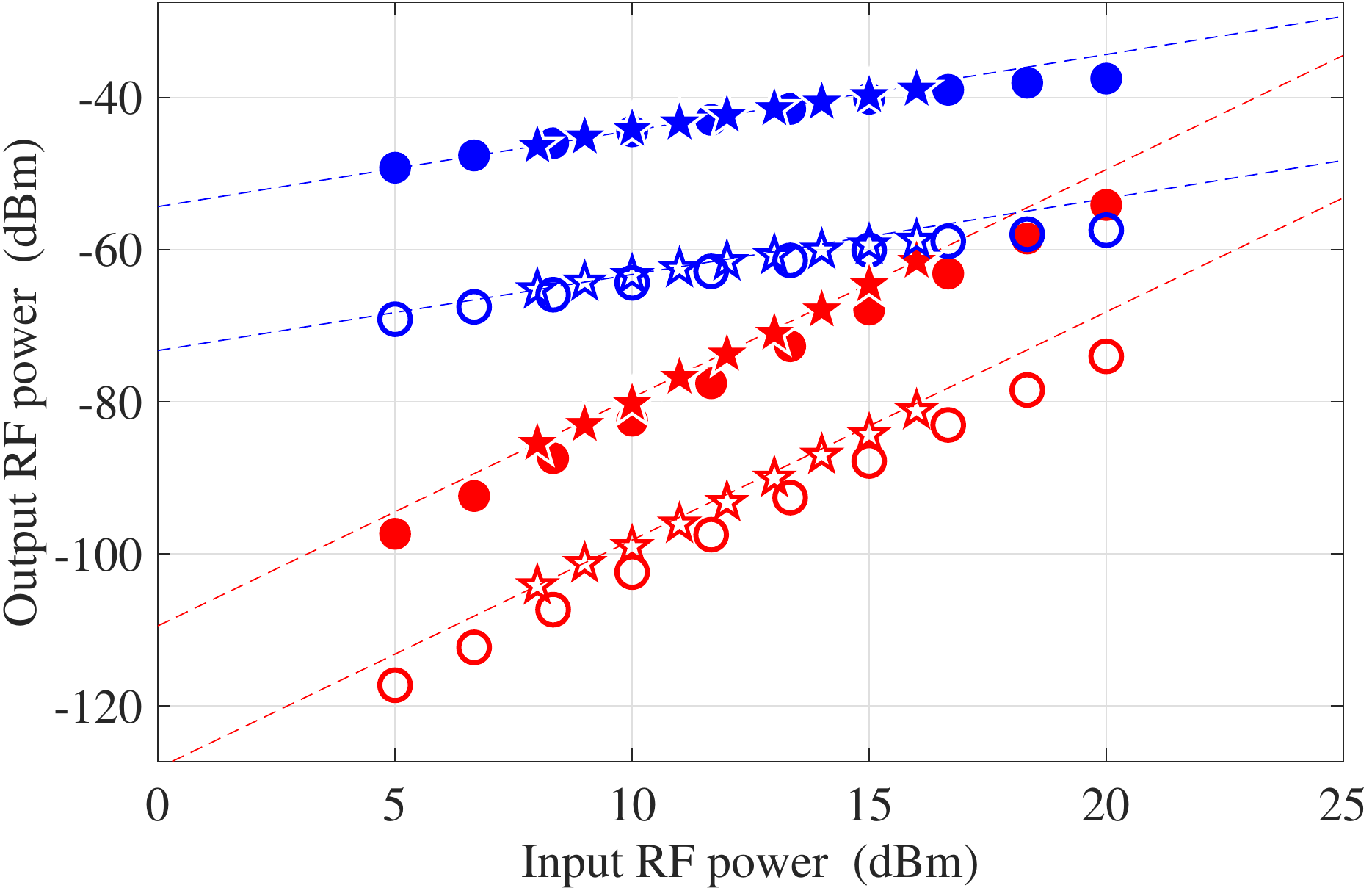}
\caption{Output fundamental (blue symbols) and IMD3 (red symbols) RF powers vs. input  RF power using  a standard MZM. Stars: measurements; circles: simulations. Open symbols: PSA off; filled symbols: PSA on. Only the signal is detected. Blue (red) dashed lines: slope 1 (3). Parameters : input pump power: 22\,dBm; input signal and idler power: -16\,dBm; HNLF length: 1000\,m.}\label{modelvalidate}
\end{figure}
Figure\,\ref{modelvalidate} compares the results obtained from the simulation with the experimental results.  The output fundamental RF powers are reproduced as blue symbols while the IMD3 RF powers are plotted in red. The numerical results correspond to the circles and the measurements to the stars. Finally, the open symbols correspond to the situation where the PSA is off (no injected pump) while the filled symbols hold for results obtained when the PSA is on (22 dBm input pump power) and provides an optical gain of 10\,dB.

First of all, the open symbols in this figure illustrate the well known fact that the nonlinearity of a standard MZM induces strong IMD3's, as already mentionned in Section \ref{sec2.E}. Second, by comparing the blue symbols with the red symbols, one can see that all the RF tones, both fundamental and IMD3's, experience a 20\,dB gain, i. e. twice the optical gain, when the PSA is ON. This shows that the PSA does not add any RF signal nonlinearity  to the one that was already created by the MZM: it just amplifies them with the same gain as the fundamentals. 

Finally, the most important feature of Fig.\ref{modelvalidate} is that the agreement between theory and experiments is very good. This means that the model developed above is valid and that we can use it to predict the behavior of a PSA in more exotic configurations.

%From Figure \ref{modelvalidate} we see that both the fundamental RF and IMD3(RF) waves attain a gain of 20 dB when the PSA is turned on in the link. This gain does not depend on the input RF power. Also, for higher values of input RF power, we see from the numerical data points, that the output RF power of fundamental(RF) and IMD3(RF) starts deviating from the lines with slopes 1 and 3 respectively. This shows us that at higher input RF powers, the MZM starts getting saturated. Also, most importantly, from this graph we see that the results from the numerical model agrees with the experimentally obtained data within about 2 dB offset error. Hence the developed numerical model is validated.

\section{Saturation Behaviour : Standard Intensity Modulator}\label{saturationRM}
\begin{figure*}[htp]
\centering 
\includegraphics[width=0.6\textwidth]{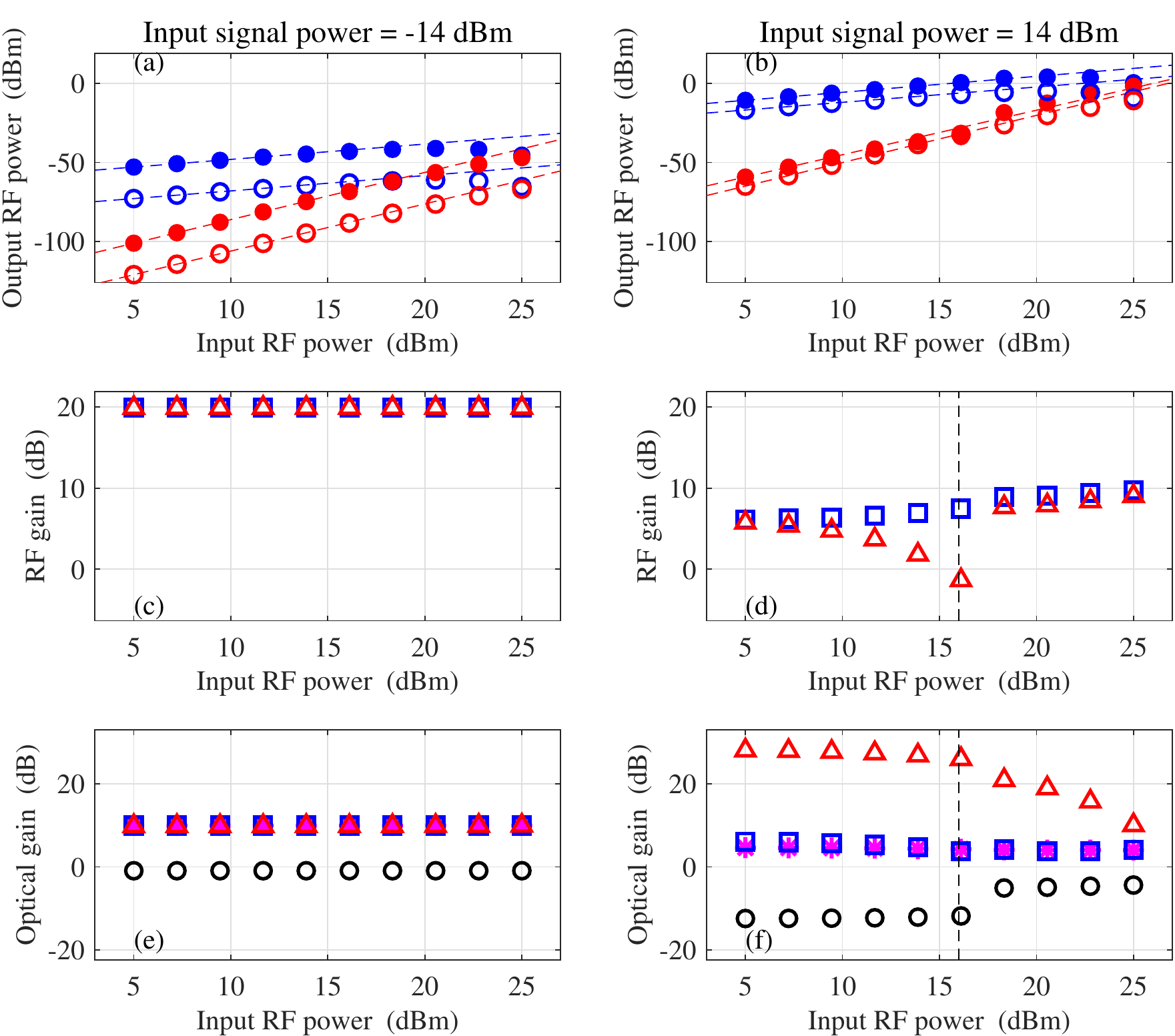}
\caption{(a,b) : Output fundamental (blue) and IMD3 (red) RF powers versus input RF power with the PSA off (open symbols) and on (full symbols). Blue and red dashed lines: guides to the eye to stress gain saturation. (c,d) :  Fundamental (blue squares) and IMD3 (red triangles) RF gains versus input RF power. (e,f) : Optical gain of the signal (magenta stars), pump (black circles), fundamental (blue squares) and IMD3 (red triangles)  tones versus input RF power. A standard MZM is used and only the signal  is detected. Signal power at the MZM input: -14\,dBm (a,c,e) and 14\,dBm (b,d,f); input pump power: 22\,dBm; HNLF length: 1000\,m.}
\label{outrfvsinrf_2sp}
\end{figure*}

In  the  previous section, the input signal and idler powers  (-16\,dBm each) were small enough so that, even after having propagated through the HNLF and experienced the PSA gain (10\,dB), their output power would remain much smaller than the pump power (22\,dBm). Then, in these conditions, one can neglect the depletion of the pump, i. e., saturation of the PSA. On the contrary, in the present section, we use our numerical model to gauge the situations where saturation of the PSA gain occurs, in order to explore whether the RF nonlinearity of the PSA increases in the presence of gain saturation. To this aim, we first observe the PSA behavior by scanning the input RF power for several input signal and idler powers (Subsection \ref{inRFstdmzmsec}) and by scanning the input signal and idler powers till gain saturation occurs (Subsection \ref{insigstdmzmsec}).

\subsection{PSA nonlinearity Evolution versus Input RF Power}\label{inRFstdmzmsec}
We can see from Eq.\,(\ref{mzmeouteq}) that, when the input RF power is increased, the powers of the higher order harmonics increase, making the modulator more and more nonlinear. We thus plot the output power of the RF fundamental (blue symbols) and IMD3 (red symbols) when the PSA is off (hollow symbols) and on (filled symbols) in Figs.\,\ref{outrfvsinrf_2sp}(a,b). In the simulations of Fig.\,\ref{outrfvsinrf_2sp} and all the following, we suppose that the detector is perfect, and that the signal undergoes 0.9 dB losses in the HNLF and 3 dB losses between the HNLF and the detector. Figure \,\ref{outrfvsinrf_2sp}(a) corresponds to 
a low (-14\,dBm) signal and idler power, including all their sidebands, at the the input of the HNLF, while in Fig.\,\ref{outrfvsinrf_2sp}(b) corresponds to a stronger (14\,dBm) input signal and idler power. The corresponding RF gains for the fundamental (blue squares) and IMD3 (red triangles) tones, when the PSA is tuned at its maximum gain, are shown in Figs.\,\ref{outrfvsinrf_2sp}(c) and \ref{outrfvsinrf_2sp}(d), respectively. Finally, Figs.\,\ref{outrfvsinrf_2sp}(e) and \ref{outrfvsinrf_2sp}(f) show the corresponding PSA optical gain for the signal (magenta stars), the pump (black circles), the optical fundamental (blue squares), and the optical IMD3 (red triangles). For each case, the relative phase between the pump, signal and the idler is adjusted to maximize the PSA gain. The other parameters are the same as in Section \ref{experiment}.

From these results, we see that for a low input signal power (-14 dBm, left column in Fig.\,\ref{outrfvsinrf_2sp}), the RF gain of fundamental and IMD3 tones is independent on the input RF power. This is due to the fact that the powers of the signal and idler  and all their sidebands are too small to saturate the PSA gain, and thus experience the same optical gain. For high RF powers ($>15\,\mathrm{dBm}$), the output RF powers of the RF fundamental and IMD3 shows some saturation due to the nonlinearity of the MZM.   

\begin{figure}[htp]
\centering 
\includegraphics[width=0.9\columnwidth]{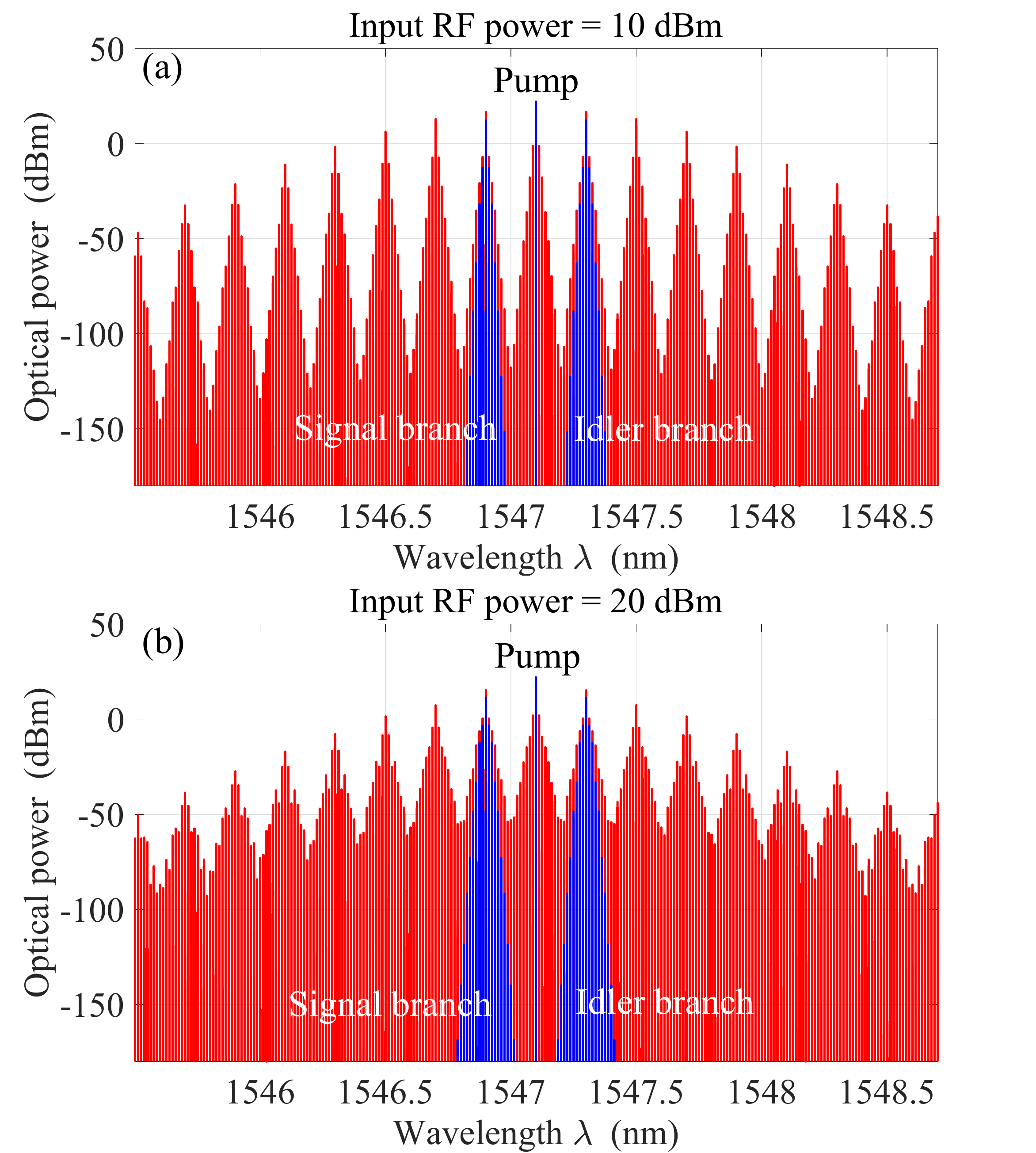}
\caption{Optical spectra at the input (blue) and output (red) of the HNLF for (a) a low  (10\,dBm) input RF power and (b) a high (20\,dBm) input RF power. Other parameters : input signal and idler power: 14\,dBm; input pump power: 22\,dBm; HNLF length: 1000\,m. Optical losses: 3.9~dB. The detector is supposed to be ideal.}
\label{inoutspectrum}
\end{figure}

On the contrary, in the case of a higher input signal power (14 dBm, right column in Fig.\,\ref{outrfvsinrf_2sp}), gain saturation occurs, as can be seen by comparing the RF gains for the fundamentals and the IMD3's in Figs.\,\ref{outrfvsinrf_2sp}(c) and (d). For input RF powers typically larger than 16 dBm (as indicated by the dashed line in Figs.\,\ref{outrfvsinrf_2sp}(d,f), a large number of strong sidebands are generated by the MZM. Then, the numerous four-wave mixing processes in the PSA modify the powers of all these frequencies in a complex manner. As an illustration, the optical spectra at the  input and output of the PSA are shown in Figs.\,\ref{inoutspectrum}(a) and (b) for relatively low (10\,dBm) and high (20\,dBm) RF powers, respectively. In Fig.\,\ref{inoutspectrum}(b), the strong input RF power leads to the fact that the beatnote between many different waves contribute to a strong IMD3 RF tone. The dip in the IMD3 RF gain in Fig.\,\ref{outrfvsinrf_2sp}(d) can be attributed to the fact that the different beatnotes contributing to the RF IMD3 are out of phase. Moreover, we also checked in this case that this result does not critically depend on the values of the modulation frequencies $\Omega_1$ and $\Omega_2$. 

Additionally, Fig.\,\ref{outrfvsinrf_2sp}(f) exhibits a decrease in the optical gain of the optical IMD3 above an input RF power of 16 dBm. This decrease is due to the presence of a higher number of strong optical sidebands at the input of the PSA, which deplete the available gain. Such a decrease of the IMD3 RF gain much below the RF gain of the fundamental tone opens interesting perspectives to use the PSA to decrease the nonlinearities of microwave photonics links, as also discussed in Refs.\, \cite{bhatia2014linearization, li2014optical}.

\subsection{Calculation of the SFDR of the link}
\begin{figure}[htp]
\centering 
\includegraphics[width=0.9\columnwidth]{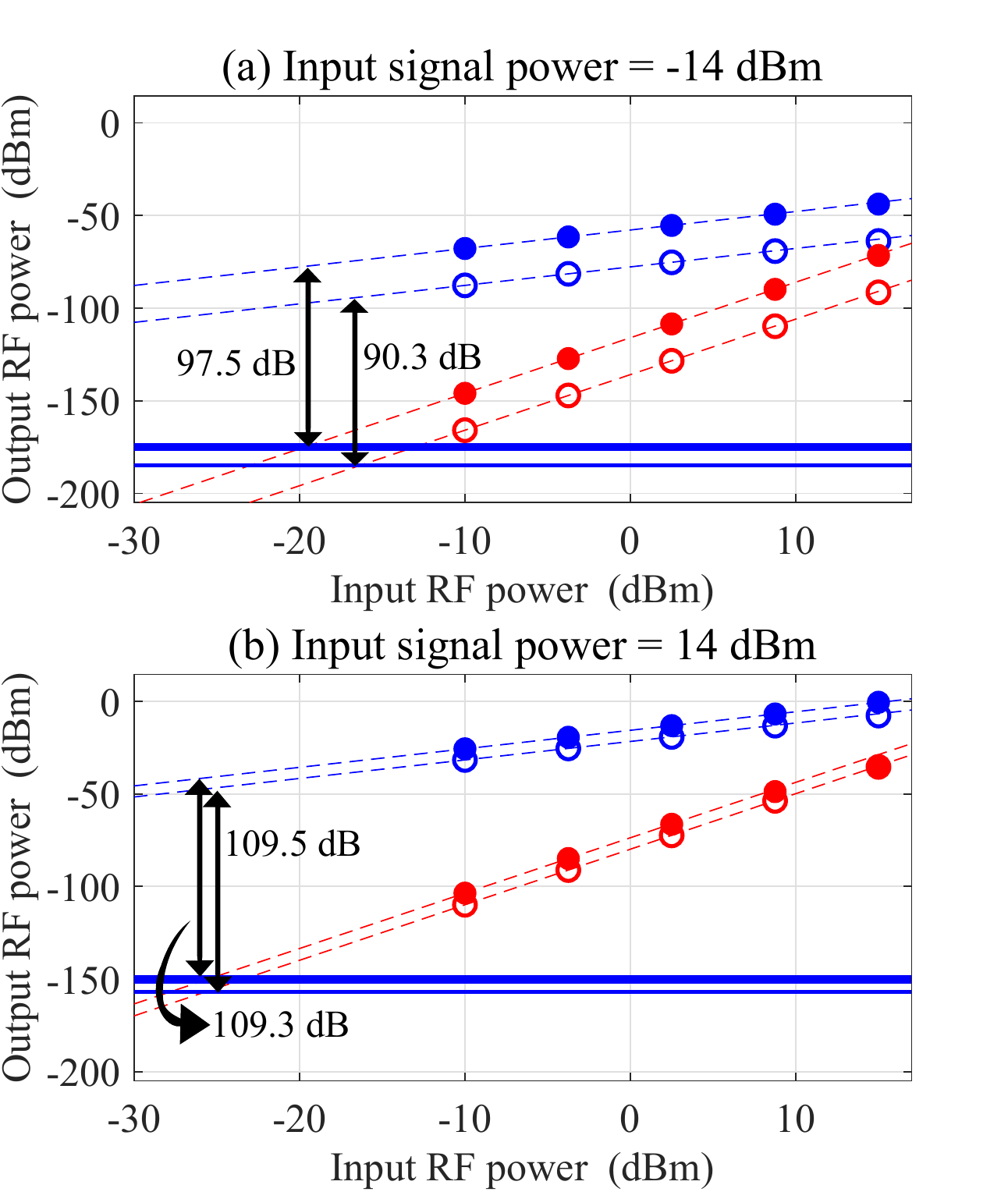}
\caption{Characterization of the SFDR of the link. The parameters of the plots (a) and (b) are the same as in Figs.\,\ref{outrfvsinrf_2sp}(a) and (b), respectively. Noise floor: shot noise level for the PSA on (thick blue line) and off (thin blue line). The corresponding SFDR's are shown in black arrows.}
\label{sfdr}
\end{figure}

In this subsection, we use the results of Fig.\,\ref{outrfvsinrf_2sp} to predict the achievable SFDR of a link using a standard MZM and such a PSA. All the parameters are the same as in Fig.\,\ref{outrfvsinrf_2sp}. Figures \ref{sfdr}(a) and \ref{sfdr}(b) show the evolution the fundamental and IMD3 RF powers as a function of the input RF power in the conditions of Figs.\,\ref{outrfvsinrf_2sp}(a) and \ref{outrfvsinrf_2sp}(b), respectively, when the PSA is on and off. Moreover, these figures also show the noise floors, supposed to be limited by the shot noise corresponding to the photo-current generated by the detector. The vertical arrows in these plots permit to extract the SFDR of the link.

One can see from these plots that the link behaves differently, in terms of SFDR, depending on whether the PSA is saturated or not. Let us first focus on the case where the optical power incident on the PSA (-14~dBm) is weak enough to keep the gain unsaturated (Fig.\,\ref{sfdr}(a)). With the PSA off, taking into account the 0.9~dB losses of the fiber and the 3~dB losses supposed to exist between the fiber and the detector, the detected optical power is -17.9~dBm, leading to a shot noise level of -184.9~dBm/Hz. When the PSA is on, it amplifies all the tones by a 10~dB optical gain. The detected optical power is then -7.9~dBm, leading to a shot noise level equal to -174.9~dBm/Hz. Besides, the 10 dB optical gain results in a 20 dB RF gain for the fundamentals and the IMD3's. The SFDR of the link is then improved by about 7~dB by the PSA. On the contrary, when the optical power incident on the PSA is equal to 14~dB, as in Fig.\,\ref{sfdr}(b), the saturated optical gain of the signal is then only 6.7~dB, increasing the shot noise level from -156.9~dBm/Hz with the PSA off to -150.2~dB/Hz with the PSA on. As can be seen from Fig.\,\ref{outrfvsinrf_2sp}(d), the RF gain for the fundamental and the IMD3's are also of the order of 7~dB in this region. Consequently, as can be seen from Fig.\,\ref{sfdr}(b), the SFDR, which is equal to 109.3~dB with the PSA off, is neither improved nor degraded by the PSA.

\subsection{PSA nonlinearity Evolution versus Input Signal Power}\label{insigstdmzmsec}
\begin{figure*}[htp]
\centering 
\includegraphics[width=0.6\textwidth]{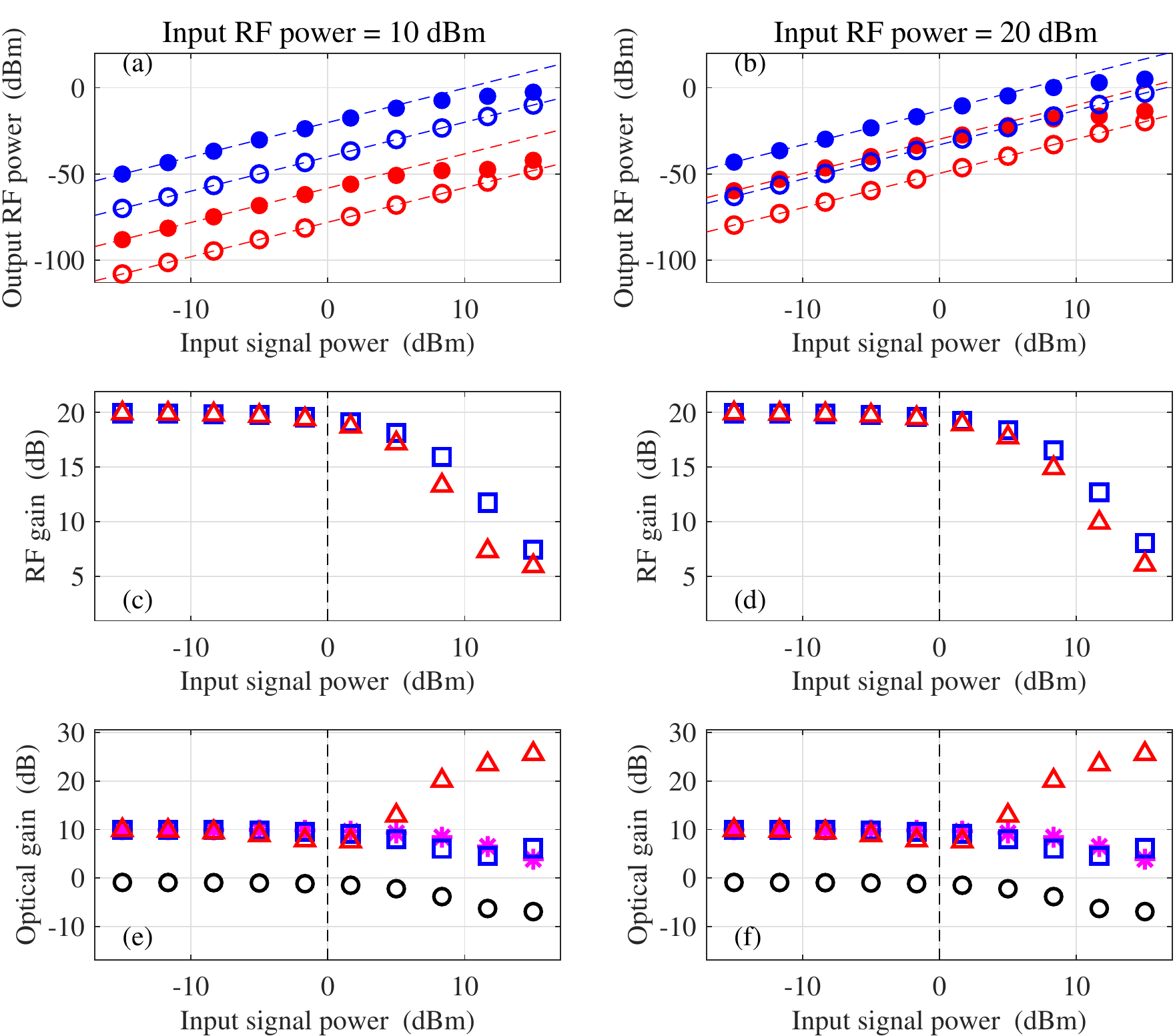}
\caption{Same as Fig.\,\ref{outrfvsinrf_2sp} except the plots are shown as a function of the input signal and idler powers, for two values of the RF input power:  10\,dBm (left column) and 20\,dBm (right column).}
\label{outrfvsinsig_2mp}
\end{figure*}
As we have seen in Fig.\,\ref{outrfvsinrf_2sp},  the increase of the input signal power leads to PSA gain saturation. We thus plot in Fig.\,\ref{outrfvsinsig_2mp} the evolution of the PSA behavior as a function of the signal and idler powers, in a manner similar to Fig.\,\ref{outrfvsinrf_2sp}. The two columns of Fig.\,\ref{outrfvsinsig_2mp} correspond to two values of the input RF power: 10\,dBm (left) and 20\,dBm (right).

%the output RF power of fundamental(RF) (blue) and IMD3(RF) (red) when the PSA is off (hollow symbols) and on (filled symbols) as a function of the input signal power in Figures \ref{outrfvsinsig_2mp} (a) and (b). In Figures \ref{outrfvsinsig_2mp} (c) and (d) we show the corresponding RF gains of fundamental(RF) (blue squares) and IMD3(RF) (red triangles) waves due to the PSA. Figures \ref{outrfvsinsig_2mp} (e) and (f) show the PSA gains of the optical waves, the signal (magenta star), pump (black circle), fundamental(optical) (blue squares) and IMD3(optical) (red triangles) as a function of the input signal power. We have used two different input RF powers : 10 dBm (for Figure \ref{outrfvsinsig_2mp} (a), (c) and (e)) and 20 dBm (for Figure \ref{outrfvsinsig_2mp} (b), (d) and (f)) for the plots. For each case, the relative phase between the pump, signal and the idler was adjusted to maximize the PSA gain of the signal. The other parameters of the simulation are same as described in Section \ref{experiment}.

Figures\,\ref{outrfvsinsig_2mp}(e) and \ref{outrfvsinsig_2mp}(f) show that the pump depletion becomes significant, i. e., the pump gain becomes much smaller than 0\,dB, for input signal powers larger than 0 dBm. This also corresponds to a decrease of the gain experienced by the signal and the optical fundamental tones. However, surprisingly, this corresponds to a strong increase, from 10\,dB up to 25\,dB, of the optical gain for the optical  IMD3 frequencies. This can be explained by the fact that, for high enough input signal powers, the power of the optical fundamental wave entering the HNLF is also quite large. Moreover, these tones become even more powerful when they travel through the fiber.  Consequently, the powers of the pump, signal, and optical fundamental become large enough to lead to a significant amplification of the optical IMD3 by four-wave mixing among these four waves, thus enhancing the optical IMD3 gain. However, as can be seen  from Figs.\,\ref{outrfvsinsig_2mp}(c) and \ref{outrfvsinsig_2mp}(d), this gain  enhancement for the optical IMD3 tones does not lead to an increase of the RF gain of the RF IMD3 tones, which actually strongly decreases for input signal powers larger than 0\,dBm. This indicates that the increase of the power of the optical IMD3 wave is counter-balanced by the influence of other PSA-generated sidebands that beat at the IMD3 RF frequency and subtract from the contribution of the beating of the signal and the optical IMD3 in the output RF IMD3 power.

To summarize, we have seen in this section that, in the case of an AM RF signal applied to the signal and idler by a standard MZM, the PSA at its maximum gain is not at all detrimental to the linearity of the carried RF signal. On the one  hand, in the absence of saturation, i.e., when the pump depletion is negligible, the RF IMD3's are just amplified by the PSA with the same gain  as the RF fundamental tones. The contribution of the  PSA to the RF nonlinearity of the link is thus completely negligible compared to the nonlinearity of the MZM. On the other hand, the saturation of the PSA can lead to a lower gain for the  RF IMD3's than for the RF fundamentals, thus leading to a partial mitigation  of  the RF nonlinearities of the MZM.

\section{Linearized Intensity Modulator}\label{saturationIM}
\begin{figure*}[htp]
\centering 
\includegraphics[width=0.6\textwidth]{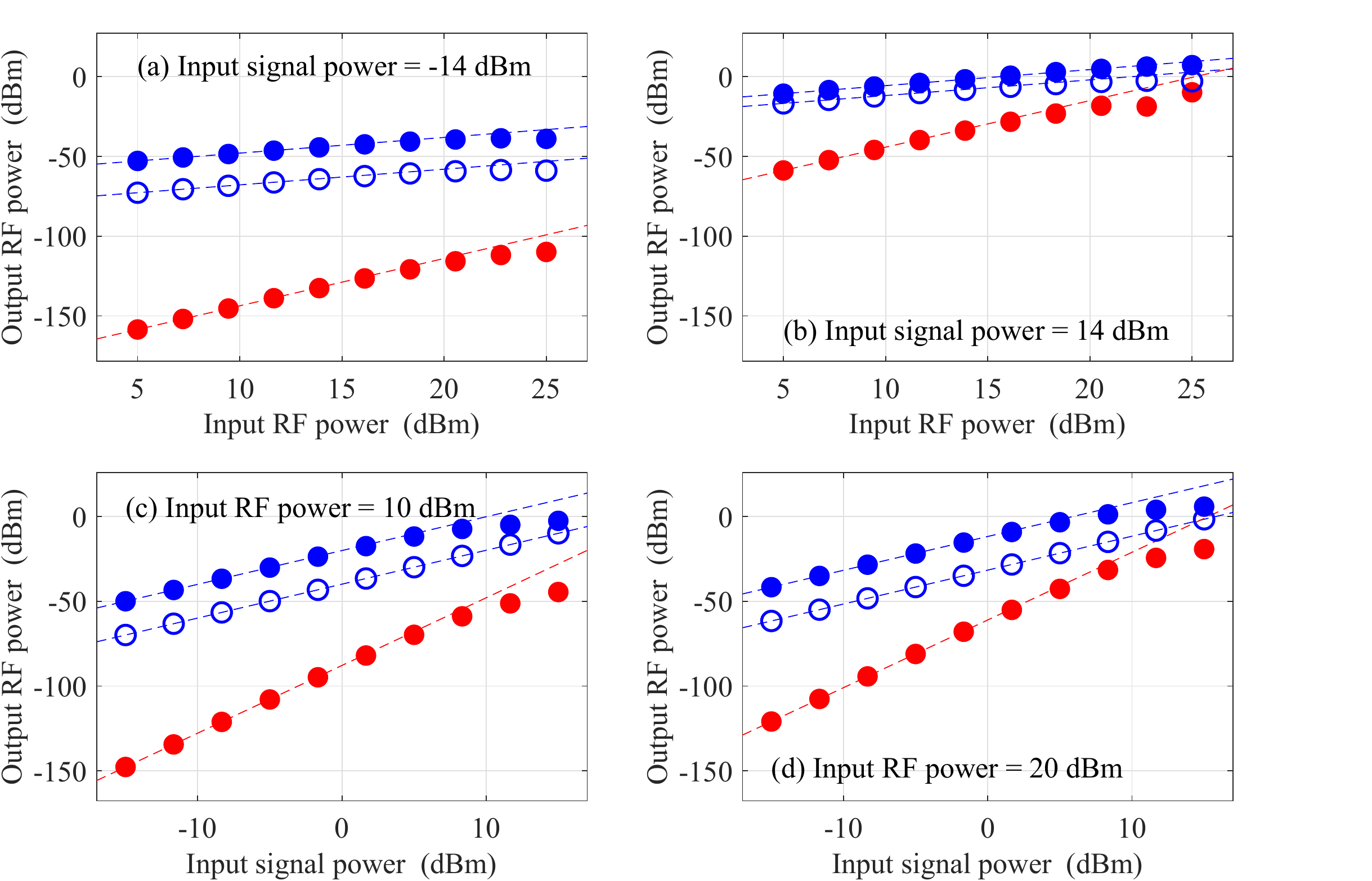}
\caption{PSA nonlinearity  with a linearized intensity  modulator. (a,b) Output fundamental (blue circles) and IMD3 (red circles) RF powers when the PSA is off (open symbols) and on (filled symbols) versus input RF power. Input signal power: (a) -14\,dBm), (b) 14\,dBm. (c,d) Same as (a,b) versus input signal power. Input RF power: (c) 10\,dBm), (d) 20\,dBm. The dashed red and blue lines are guides to the eye.}
\label{linmodgrph}
\end{figure*}
\begin{figure}[htp]
\centering 
\includegraphics[width=0.95\columnwidth]{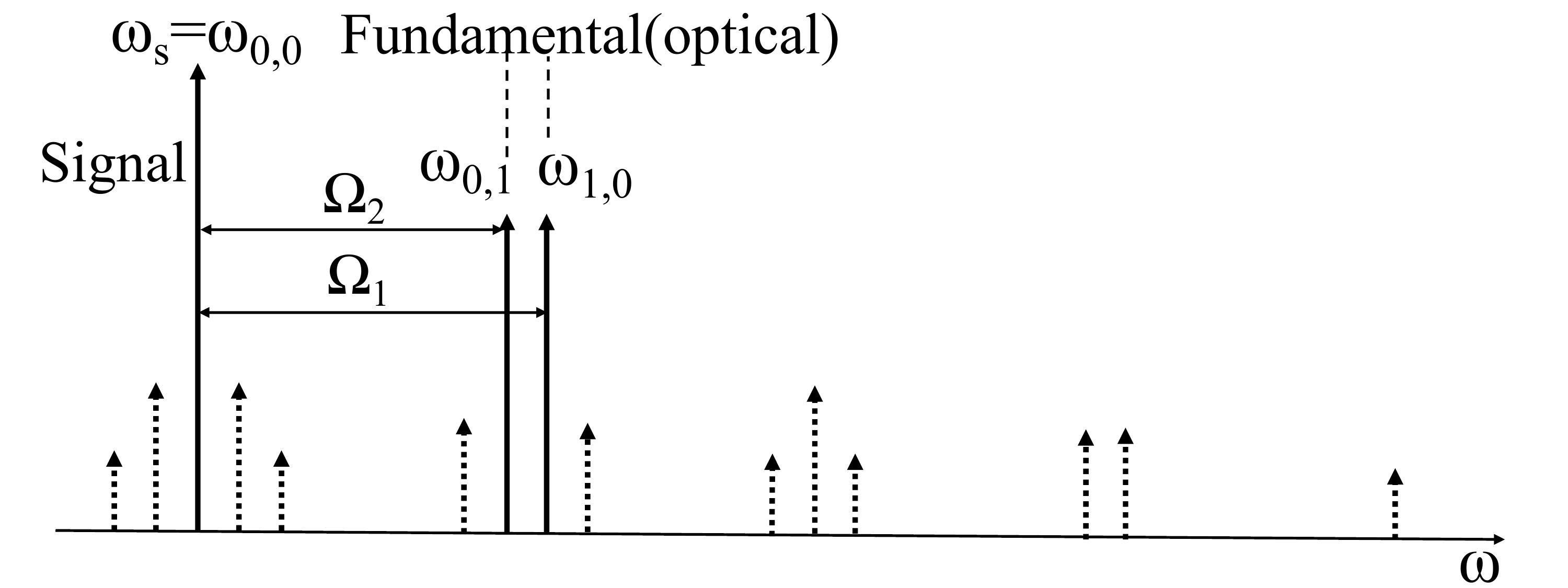}
\caption{Solid arrows: optical spectrum at the output of a perfectly linear intensity modulator fed with two modulation frequencies $\Omega_1$ and $\Omega_2$. Dashed arrows: waves that would have been generated by a standard MZM (not to scale).}
\label{linmodspectrum}
\end{figure}
We have seen in the preceding sections, the nonlinearities created by a standard MZM largely exceed those which are induced by the nonlinearities of the PSA. Consequently, in order to determine the RF nonlinearities created by the PSA itself, we consider in the present section the case where the RF modulation is applied to the optical carriers using a perfectly linear intensity modulator. Such a modulator would generate no high order sidebands while modulating the signal. Hence, only the  first order sidebands are supposed to be present, as shown in Fig.\,\ref{linmodspectrum} in the case of modulation by the two frequencies  $\Omega_1$ and $\Omega_2$. In such a case, the RF IMD3's detected at the output is solely due to the PSA.

Figure\,\ref{linmodgrph} shows the simulation results obtained in this case. In the case of a relatively low input signal and idler power (see Fig.\,\ref{linmodgrph}(a)), we can see that the RF IMD3's created by the PSA have an extremely small power level. For example, compared with the case of a standard MZM (see Fig.\,\ref{outrfvsinrf_2sp}(a)), the IMD3 power obtained in Fig.\,\ref{linmodgrph}(a) is reduced by about 58\,dB when the input RF power is equal to 10\,dBm. 

However, when gain saturation enters the picture, the optical nonlinearity of  the fiber becomes strong enough to create significant nonlinearities for the RF signal. This is particular visible in Figs.\,\ref{linmodgrph}(b) and \,\ref{linmodgrph}(d), for relatively high input optical and RF powers. In such cases, one can see that the severe reduction by saturation of the gain of the fundamental RF tone is accompanied by an efficient generation of IMD3's. The powers of the created RF IMD3's can indeed become almost as large as those of the RF fundamentals.

\begin{figure}[htp]
\centering 
\includegraphics[width=1\columnwidth]{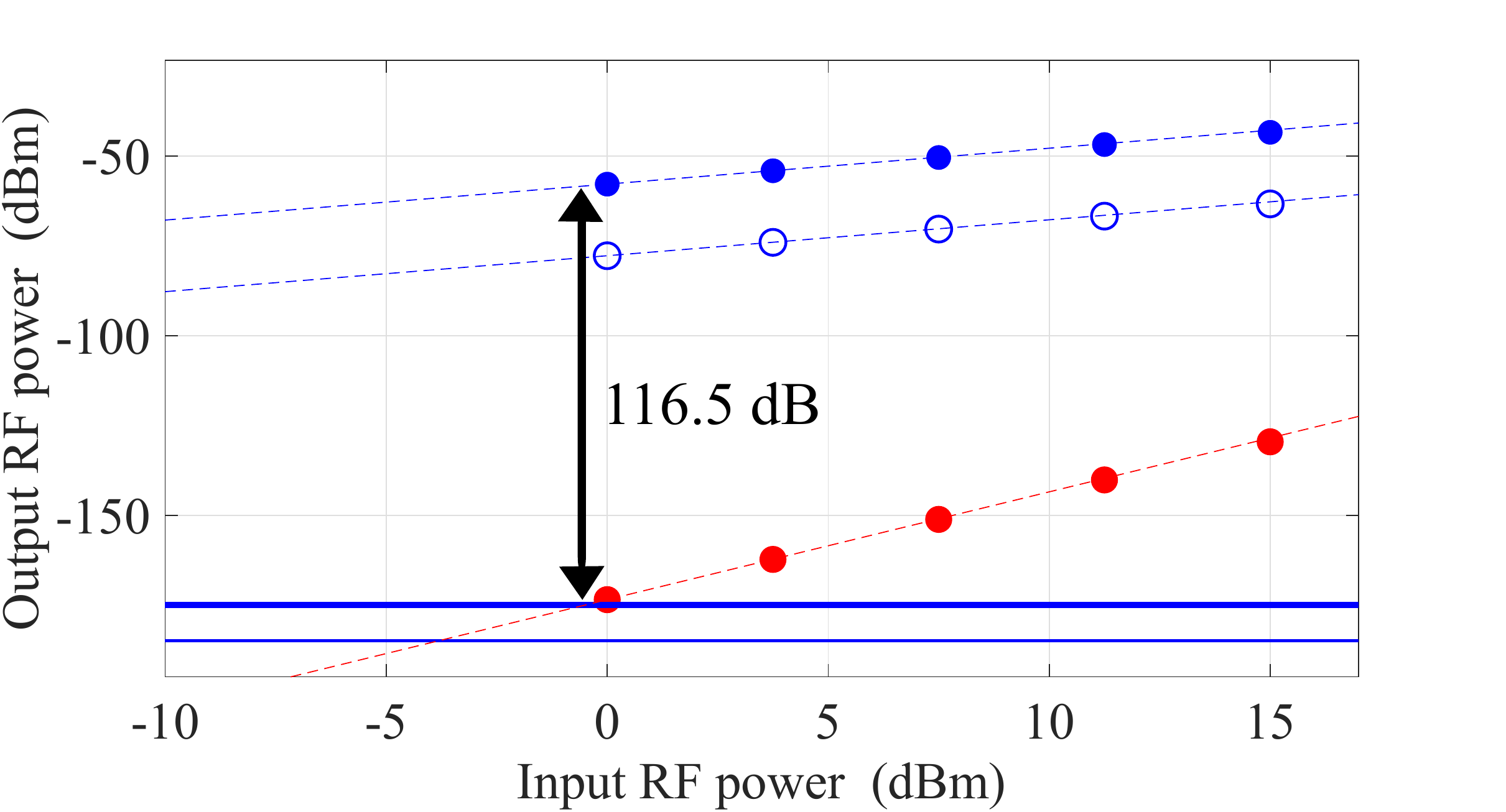}
\caption{Same as Fig.\,\ref{sfdr}(a) with the parameters of Fig.\,\ref{linmodgrph}(a). The input signal power is -14~dBm.}
\label{sfdrlinmod}
\end{figure}

It is clear from the plots of Fig.\,\ref{linmodgrph} that the most interesting situation from the point of view of applications is the one of Fig.\,\ref{linmodgrph}(a), where a linearized intensity modulator is used and the PSA gain is not saturated. Indeed, in these cases, the IMD3's at the output of the PSA are very small, much smaller than in the case where a standard modulator is used (see Fig.\,\ref{outrfvsinrf_2sp}(a)). We thus choose to investigate the SFDR of the link in the case of Fig.\,\ref{linmodgrph}(a). The result is reproduced in Fig.\,\ref{sfdrlinmod}. With the PSA on, the -14~dBm signal power incident on the PSA is amplified by 10~dB. After the 0.9~dB losses of the fiber and 3~dB extra losses, -7.9~dBm optical power incident on the detector leads to a shot noise level of -174.9~dBm/Hz. The arrow in Fig.\,\ref{sfdrlinmod} then shows that the SFDR is then equal to 116.5~dB. By comparison with Fig.\,\ref{sfdr}(a), this corresponds to a 19 dB improvement. One interesting point here is that the SFDR achieved with the linearized modulator and the PSA with a relatively small input signal power (-14~dBm) is larger than the one obtained with a standard modulator with a much larger input signal power (14 dBm).

\section{Discussion and Conclusion}\label{discussion}
In this paper, we have discussed how the insertion of a PSA inside a microwave photonics link based on amplitude modulation alters the RF nonlinearities of the link. To this aim we used a numerical model to evaluate the powers of the fundamental and IMD3 RF waves at the output of the link. We validated the model by comparison with experimental results. 

Different situations were considered, based either on a standard Mach-Zehnder modulator or a linearized modulator to transfer the RF signal to the optical carriers. We have also compared the cases of weak and strong optical powers and low and strong RF powers applied to the modulator. However, in this analysis we have not considered the random fluctuations of the ZDW of the fiber that can often degrade the noise and amplification performance of the link \cite{velanas2006impact,karlsson1998four}.

The first conclusion of this paper is that, in spite of its intrinsic optical nonlinearity, a PSA used at its maximum gain can behave in a surprisingly linear manner from the point of view of the carried RF signal. This linear behavior is observed when the input signal and idler powers are weak enough for the pump depletion to be negligible. Then, we have seen that, in the case of a standard MZM, the RF nonlinearity of the PSA is completely negligible compared to the one of the modulator. Even if one removes these latter nonlinearities by using a perfectly linear amplitude modulator, the nonlinearities induced by the PSA remain extremely small as long as the gain is not saturated.

In terms of SFDR, the conclusions of this work are the following. On the one hand, when one uses a standard MZM, the use of a PSA improves the SFDR when the gain is not saturated, while it neither improves or degrades it when the gain is saturated. On the other hand, in case one uses a perfectly linear modulator, the RF nonlinearities created by the PSA itself are so small that the SFDR of the link can be significantly larger than in the case of a standard MZM. This can be useful to avoid the use of high signal powers or in the case the detector itself cannot handle such high powers before becoming nonlinear.

The picture becomes drastically different when the input signal power is strong enough to induce gain saturation via pump depletion. But, in this case, the PSA behavior is different in the case of a standard MZM and in the case of a linearized modulator. In the first case, we have shown that the strong signal nonlinearities created by the intrinsically nonlinear MZM can actually be strongly reduced by the PSA. The amplifier can thus be used to mitigate the link nonlinearities and improve its dynamic range. On the contrary, when one uses a linearized amplitude modulator, the saturated PSA can create signal nonlinearities that can become extremely detrimental to the linearity of the link. Indeed, we have seen that in some situations, extreme gain saturation can generate RF IMD3's with powers as large as those of the RF fundamentals.

%Then using the numerical model, we investigated the performance of the link under high modulation power and high input signal powers. In the case of using higher modulation power, a large number of strong sidebands are generated that when amplified by the PSA, leads to a lower RF gain for the link. Also, when we increase the input signal power, above a certain threshold, the RF gain of the fundamental(RF) and IMD3(RF) waves start decreasing due to the effect of pump depletion and the presence of large number of amplified sidebands. Using the case of a perfectly linear modulator, we also found that the distortions induced by the PSA process are highly sensitive to the input signal power of the link. When the pump remains undepleted or the input signal power is low, the modulator is the main source of distortion, while for higher signal powers, the PSA becomes dominant.  

In this work we have considered only a direct detection scheme to detect the optical signal after amplification. However, a coherent detection scheme can be used to further improve the noise performance of the link \cite{agarwal2011optically}. Also, recent developments suggest the use of few mode fibers (FMF) in a microwave photonic link to increase the link's optical power handling capabilities \cite{wen2017few}. One may wonder how incorporation of a PSA in a FMF based microwave photonic link could be achieved for further performance improvement. In particular, the problem of the control of the relative phases between the pump(s) and the different modes seems particularly tricky. Moreover, this work also opens new perspectives for the investigation of link nonlinearities using fibers exhibiting dispersion fluctuations \cite{nissim2014performance,bagheri2016gain,gao2017theoretical} and dispersion oscillations \cite{mussot2018modulation}.

\section*{Acknowledgement}
We thank the anonymous reviewers for their critical suggestions to improve the quality of this paper.

\section*{Disclosures}
The authors declare no conflicts of interest.
%\section{Acknowledgement}
%This work was supported by grants from the CNRS and ENS Paris-Saclay, France. 

% Bibliography
\bibliography{Biblio_Resubmitted}

% Full bibliography added automatically for Optics Letters submissions; the following line will simply be ignored if submitting to other journals.
% Note that this extra page will not count against page length
%\bibliographyfullrefs{Biblio_Resubmitted}

%Manual citation list
%\begin{thebibliography}{1}
%\bibitem{Zhang:14}
%Y.~Zhang, S.~Qiao, L.~Sun, Q.~W. Shi, W.~Huang, %L.~Li, and Z.~Yang,
 % \enquote{Photoinduced active terahertz metamaterials with nanostructured
  %vanadium dioxide film deposited by sol-gel method,} Opt. Express \textbf{22},
  %11070--11078 (2014).
%\end{thebibliography}

% Please include bios and photos of all authors for aop articles
%\ifthenelse{\equal{\journalref}{aop}}{%
%\section*{Author Biographies}
%\begingroup
%\setlength\intextsep{0pt}
%\begin{minipage}[t][6.3cm][t]{1.0\textwidth} % Adjust height [6.3cm] as required for separation of bio photos.
%  \begin{wrapfigure}{L}{0.25\textwidth}
%    \includegraphics[width=0.25\textwidth]{john_smith.eps}
%  \end{wrapfigure}
%  \noindent
%  {\bfseries John Smith} received his BSc (Mathematics) in 2000 from The University of Maryland. His research interests include lasers and optics.
%\end{minipage}
%\begin{minipage}{1.0\textwidth}
%  \begin{wrapfigure}{L}{0.25\textwidth}
%    \includegraphics[width=0.25\textwidth]{alice_smith.eps}
%  \end{wrapfigure}
%  \noindent
%  {\bfseries Alice Smith} also received her BSc (Mathematics) in 2000 from The University of Maryland. Her research interests also include lasers and optics.
%\end{minipage}
%\endgroup
%}{}

\end{document}